# Defining a Lingua Franca to Open the Black Box of a Naïve Bayes Recommender

Kenneth L. Hess*, Hugo D. Paz
Science Buddies, Milpitas, CA

**Abstract**

Many AI systems have a black box nature that makes it difficult to understand how they make their recommendations. This can be unsettling, as the designer cannot be certain how the system will respond to novelty. To penetrate our Naïve Bayes recommender's black box, we first asked, what do we want to know from our system, and how can it be obtained? The answers led us to recursively define a common lexicon with the AI, a lingua franca, using the very items that the system ranks to create meta-symbols recognized by the system, and enabling us to understand the system's knowledge in plain terms and at different levels of abstraction. As one bonus, using its existing knowledge, the lingua franca can enable the system to extend recommendations to related, but entirely new areas, ameliorating the cold start problem. We also supplement the lingua franca with techniques for visualizing the system's knowledge state, develop metrics for evaluating the meaningfulness of terms in the lingua franca, and generalize the requirements for developing a similar lingua franca in other applications.

**Recommender Systems**

Recommender systems are used in a variety of applications ranging from those that recommend products for purchase to those that recommend movies to watch or people to date. [1] [2] [3] [4] [5] Our recommender, the Topic Selection Wizard (www.sciencebuddies.org/tsw), makes personalized suggestions of hands-on science projects that K-12 students will find intrinsically interesting, greatly increasing the odds that the students will have a positive learning experience. [6] First available in 2002, approximately 10 million students have used increasingly sophisticated versions of the Wizard.

The Wizard is a model-based recommender, with separate phases for training (or model-building) and prediction. [7] We prefer to label the Wizard's training component as its *historical* process, because it looks backwards and simply records what occurred. The process records vectors comprised of a user action or actions and a subsequent user choice. In the Topic Selection Wizard, we ask every user the same 26 questions about their everyday interests, such as, "Is math your favorite subject in school?" To every question, the user can respond "yes," "sometimes," or "no." We also track whether the same user expresses an interest in one of more than 1,200 project ideas by creating a satisfaction event (printing it out or otherwise implicitly showing satisfaction with the idea). For each project idea, our historical process periodically tabulates the vectors comprising approximately 1.5 million users' survey responses and satisfaction events if any, building the historical record as a matrix that shows the probability of creating a satisfaction event for a specific project idea while answering each question in a particular way.

Our recommender, like most other model-based recommenders, runs the historical process periodically in a batch mode, but in principle it could be updated in real time with each new user interaction.

The recommender is unusual in having complete information for each user (no matrix completion problem); however, neither this architecture nor the recommendation algorithm that follows is a requirement of the lingua franca capabilities described later in this article.

The *recommendation* process involves making a prediction about what a specific user would like. The Topic Selection Wizard recommender is a Naïve Bayes classifier. Bayes' theorem is stated mathematically as follows:

$$P(A \mid B) = \frac{P(B \mid A)P(A)}{P(B)}$$

where $A$ and $B$ are events and $P(B) \neq 0$.
- $P(A)$ and $P(B)$ are the probabilities of observing $A$ and $B$ without regard to each other.
- $P(A \mid B)$, a conditional probability, is the probability of observing event $A$ given that $B$ is true.
- $P(B \mid A)$ is the probability of observing event $B$ given that $A$ is true.

In terms of our recommender system:

$$P(Project \mid Response) = \frac{P(Response \mid Project)\, P(Project)}{P(Response)}$$

where *Project* is a project idea and *Response* is the user's response to one of our science interest survey questions.
- $P(Project)$ is the probability of a user satisfaction event (printing, emailing, marking as a favorite or viewing a project idea for an extended period of time) without regard to how the user responds to the survey question. It is also called the *prior probability*.
- $P(Response)$ is the probability of responding to the science interest survey question in a specific way.
- $P(Project \mid Response)$, a conditional probability, is the probability of a user satisfaction event for

---

* Corresponding author: ken.hess@sciencebuddies.org

*Project* given that *Response* is answered in a specific way.
- P(*Response* | *Project*) is the probability of observing the *Response* to the science interest survey question given that the user has created a satisfaction event for *Project*.

The Wizard is a Naïve Bayes classifier because we make the naïve, simplifying assumption that the user response to each question in the science interest survey is independent from the response to every other question. We can make an additional simplification because P(*Response*) is the same for every project idea. Since our purpose is to rank the different project ideas, and P(*Response*) is a constant, when building the probabilities matrix we can ignore it.

A new user activates the recommendation process by taking the science interest survey. For every project idea in the system, the system calculates the probability of the user creating a satisfaction event for the project idea (a proxy for being interested in the project idea) by multiplying the prior probability by the probability that each of the 26 question responses would result in a satisfaction event.

$$P(Project \mid All\ Responses) = P(Project) \times \prod_{q=1}^{n} P(Response \mid Project)_q$$

- P(*Project* | *All Responses*) is the probability of the user liking a Project Idea.
- $q$ is a question.
- P(*Project*) is the prior probability for the Project Idea.
- P(*Response* | *Project*)$_q$ is the probability that a question was answered a certain way for a given Project Idea.
- $n$ is the number of questions answered.

These recommendation probabilities are sorted with the highest probabilities representing the top recommendations.

In practice, we see more than three orders of magnitude difference in prior probability between the most popular and least popular project ideas, so the prior probability can totally obscure the results coming from the question responses, which have a much narrower range. Thus, from an engineering point of view, we treat the prior probability as a tunable parameter, tuned to reflect the purpose at hand. In our case, we want to expose students to ideas they may not have thought of, but would find intrinsically interesting, rather than displaying what would amount to a popularity list, so we underweight the prior probability as compared to a strict Bayesian implementation.

For our current science interest survey having 26 questions with three values each, there are $3^{26}$ possible combinations of answers = 2,541,865,828,329 possibilities or states of the system. In our production environment, we run our recommender system with the history of the most recent 1.5 million users, approximately.

Essentially no two users answer the survey questions exactly the same (the exceptions are users on a fool's errand to answer all the questions the same); even so, our history typically covers less than one millionth of the possible state space. Taking this one step further, because each individual who by expressing an interest contributes to the history for a project idea will have an infinitesimally small chance of repeating the same contribution as another individual, the history for each project idea will be spread over a wide area within the state space, making the recommender system highly probabilistic.

**The Black Box**

Many AI systems have a black box nature that makes it difficult to understand how they make their recommendations. [8] [9] This can be unsettling, especially when the stakes are high and the designer cannot be certain how the system will respond to novelty.

The very concept of a black box is fuzzy, because black boxes span a range of darkness. Whether something is a black box also depends on who is asking (and what they already know). To many users, an everyday computer application is a black box, but not to most geeks. In a true black box, the internal workings would be completely opaque to everyone, but such black boxes are rare. In many cases, the darkness of a black box can be measured by how much effort it takes to learn what you want to know. In our case, we can trace through our system's internal workings, but it's not always helpful. We know how it works, but not at the level of abstraction we want, so it's often impractical to answer important questions about its behavior. Call our recommender system dark gray.

Instead of asking, "How does the system work?" we began the process of opening our black box by asking a more specific, operational question, "What do we want to know?" For our end users, we wanted something richer than a ranked list of project ideas. Some users want a description of what is likely to interest them, rather than a list of examples. Other users would benefit from a description of their interests, simply to reinforce the list of examples. [10] [11] [12]

As designers and maintainers of the system, we want better tools to understand how changes to the system impact the quality of its recommendations. For example, if we add or subtract questions from our science interest survey, what is the relative impact on the ranking of different areas of science? Do the various disciplines rank similarly to each other after the change, or do the changes result in a skew towards one area or another.

Importantly, we also want to identify deficiencies in the breadth of our content offerings. Can the system identify areas of science that are intrinsically interesting to our audience that we do not currently offer?

All of these questions require the ability to generalize: to understand recommendations at multiple levels of abstraction. Is our recommender limited to ranking a list of project ideas, or has the interaction with millions of users given it a richer understanding of its domain? How deep does our recommender's understanding go? It was clear that to obtain this knowledge we needed the means to symbolically interact with the system, a bridge language common to us and the system, a lingua franca.

**Constructing a Lingua Franca**

To build a bridge language we looked for the greatest common factor between the recommender system and ourselves and that is the set of project ideas. The recommender system sees a project idea as an ordered bag of numbers (a 78-dimensional vector to be precise), representing the history of user interactions with the science interest survey and the project idea. To us, a project idea represents a set of instructions in an area of science, a means to excite a student about something of intrinsic interest, a possible career. We cannot truly understand the 78-dimensional vector of the recommender system and it knows nothing of the concepts we understand; however, we both know project ideas, the greatest common factor between us. We recursively construct the lingua franca using the project ideas, the very items recommended by the system. They represent the intersection of the recommender system's historical records and our human view of the world.

To construct a lingua franca, we need to look beneath the surface of the recommended items and see them as symbols. In the case of our recommender, each science project idea symbolizes a wide range of ideas, concepts, and objects, ranging from the lab techniques the project requires to the profession of individuals who typically perform similar work (see Table 1).

- The discipline of science or engineering containing the project idea
- Lab techniques used to perform the experiment
- Engineering techniques used to solve the problem
- Equipment used
- Materials used
- Location where similar work is usually carried out
- The profession of people who do the work on a regular basis
- Characteristics of people who do the work, such as their education
- Those who benefit from the work described in the project idea

**Table 1. Concepts symbolized by a project idea.**

The lingua franca construction process is primarily one of definition rather than training or learning. The recommender system history already contains the necessary knowledge. To generate a lexicon to use in communicating with our recommender, we recursively define meta-symbols as sets of project ideas or other meta-symbols.

- A project idea represents the simplest symbol in our recommender system (the base case).
- A symbol is defined in terms of other symbols (the recursion step).

For example, we can define the meta-symbol *aeronautical engineer* as the set of all symbols (project ideas) entailing the work of an aeronautical engineer. Each symbol contributes something to the meaning of the meta-symbol. Likewise, we define the history for the meta-symbol *aeronautical engineer* as the sum of the histories, excluding duplicates, for the same set of symbols and meta-symbols. Any symbol or meta-symbol can be included in an indefinite number of definitions.

Put another way, one project idea symbolizes many things; the union of many project ideas defines one, non-unique symbol:

$$\text{Project Idea}_1 \cup \text{Project Idea}_2 \cup ... \text{Project Idea}_n \rightarrow \text{Symbol}_i$$

This is quite like the process of defining words in human language. A *verbal* definition defines words in terms of other words. [13] And, like in human language, these definitions are inherently imprecise.

> ...definitions represent symbolizations of mental processes which establish more or less communicable boundaries for the signification mappings of symbols. For real people in the real world there cannot be perfect, i.e., arbitrarily precise definitions, since all communicable boundaries are fuzzy, ambiguous, and are subject, to some degree, to arbitrary interpretation. [14]

A Naïve Bayesian recommender works with probabilities rather than certainties, and we can rank meta-symbols simultaneously with symbols (project ideas) with no change to the underlying system. The recommender ranks all items, symbols, and meta-symbols at the same time all the time.

Some meta-symbols may be defined by sets of symbols that are similar or even identical to those for other meta-symbols. Fortunately, we can sometimes increase the specificity of the system by simply changing the *focus* to differentiate similar symbols. For example, within our comparatively small system, the profession of *aerodynamicist* and the tool *wind tunnel* might have similar definitions; however, the system can distinguish these two meta-symbols based on the focus: professions or tools. At the user interface level, the system can focus as appropriate on either project ideas, meta-symbols (our lexicon), or even a specific subset of meta-symbols depending on the context and user needs.

Currently, our 26-question survey and approximately 1,200 base symbols (project ideas) are supporting more than 2,000 meta-symbols, limited only by our ability to add definitions. Current meta-symbols include disciplines of science ranging across five different levels from areas like physical science on the high end to X-ray astronomy on the low end, items in the materials lists for the project ideas, scientific concepts such as Bernoulli's Principle, interdisciplinary areas of science, scientific tools, and STEM careers (see Table 2).

Given the nature of our current system, our lingua franca is just a lexicon, but that meets our immediate needs for symbolic communication.

| Symbol or Meta-symbol | Examples from the Lexicon | Description |
|---|---|---|
| Base symbols (project ideas) | (Base symbols underlie the lexicon, but they are not part of the vocabulary.) | Approximately 1,200 items, such as:<br>• Rocket Aerodynamics<br>• Effect of Friction on Objects in Motion<br>• Using a Digital Camera to Measure Skyglow<br>• Making Milk Curdle with Pineapple Enzymes<br>• Bits, Bytes, and Bases: Write a JavaScript Binary/Decimal/Hexadecimal Converter<br>• How Vines Find Their Spines: Thigmotropism in Morning Glory Tendrils |
| Major areas of science | • Behavioral & Social Science<br>• Earth & Environmental Science<br>• Engineering<br>• Life Science<br>• Math & Computer Science<br>• Physical Science | |
| Areas of science | 36 areas, such as:<br>• Aerodynamics & Hydrodynamics<br>• Astronomy<br>• Computer Science<br>• Genetics & Genomics<br>• Human Behavior<br>• Plant Biology | |
| Sub-areas of science, scientific concepts, and scientific techniques | Approximately 1,100, such as:<br>• Anatomy & body systems<br>• Gel electrophoresis<br>• Periodic motion<br>• Solubility<br>• Microbial growth<br>• Software languages<br>• Ohm's law | |
| Interdisciplinary areas of science | Approximately 630, such as:<br>• Astronomy + computer science<br>• Electricity & electronics + technology art<br>• Environmental science + sociology<br>• Genetics & genomics + space exploration<br>• Music + pure mathematics | All pairwise combinations of 36 different areas of science, some of which may be unlikely to occur in the real world, but most do. |
| Common tools and materials | Approximately 250, such as:<br>• Multimeter<br>• Magnet<br>• Pipette<br>• Petri dish<br>• Breadboard (for electrical circuits)<br>• Daphnia | |
| STEM professions and careers | Approximately 150, such as:<br>• Aerospace engineer<br>• Chemist<br>• Food science technician<br>• Neurologist<br>• Park ranger<br>• Sociologist | |

**Table 2. Examples from the lexicon.** Using subsets of the approximately 1,200 base symbols, we defined more than 2,000 meta-symbols to create the recommender's lexicon.

**Visualization**

While developing the lingua franca technique we frequently analyzed problems with informal sketches and images in the mind's eye. As our work progressed, we developed more refined visualizations that enabled us to validate key concepts of our model by asking the question, does the picture look as expected?

The history of each base symbol (project idea) in our recommender system is defined by 78 probabilities representing the three possible answers for each of the 26 questions in our science interest survey. If humans could visualize 78 dimensions, we could plot each base symbol in a 78-dimension space that would show exactly how each symbol relates to every other. Because we cannot easily visualize 78 dimensions, we applied Sammon mapping [15] to reduce those 78 dimensions to a lower number, albeit with errors. A 3-dimensional or 2-dimensional Sammon mapping can be no better than an approximation, but it allows us to plot the symbols in a space that we can easily perceive.

Sammon mapping tries to preserve the distance between points in the higher-dimensional space as it iteratively projects them to a lower-dimensional space. Information is lost in this process and the resulting projection has an arbitrary scale without units; nonetheless, because it gives us an estimate of the distance between symbols, it enables us to visualize similarity, dissimilarity, and relative position in the historical record. Symbols in close proximity have a similar history, and symbols far apart less so. Symbols at opposite poles of the entire set of symbols represent student interests that are roughly the opposite of each other in our knowledge space. Figures 1 and 2 show Sammon mappings for a selection of meta-symbols.

Because each user answers the questions that create the history, we also have the information needed to plot our users (based on their survey responses) in the same space (see Figure 3).

**Meaning**

Can definitions so fuzzy and probabilistic actually be meaningful? The Science Buddies website has a STEM career section with more than 150 careers. As a litmus test for our recursive definitions, we built a history for these native career items and compared how the recommender system ranks them against meta-symbols defined by project ideas that entail the work of the same STEM careers.

There are some limitations to this test. The native items arguably give a complete picture of a career, whereas the corresponding meta-symbols will overweight or underweight important characteristics of some careers simply because we have not developed project ideas that represent all aspects of most careers. They represent *ad hoc* and incomplete definitions that cannot be expected to map perfectly. On the other hand, we have very limited history available for the native careers, which introduces significant noise into their history and the comparison with the meta-symbols. Despite these limitations, there is an average Spearman Rank Correlation coefficient [16] of 0.60 between the native career items and the meta-symbols for a sample of 1,000 typical users, which suggests a strong relationship given the limitations. Figure 4 shows a bubble plot visualization of this correlation.

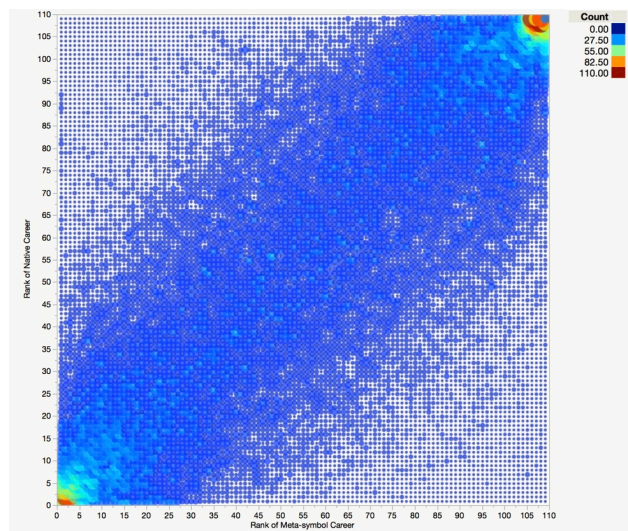

**Figure 4. Bubble plot of the ranks of 109 native career items vs. the corresponding meta-symbol career definitions, for 1,000 representative individuals.** Meta-symbol career definitions with fewer than four underlying base symbols have been filtered out. This comparison has an average Spearman Rank Correlation coefficient of 0.60 for all users' rank correlations.

The greatest value of defining meta-symbols accrues when we don't have a corresponding, native item, so metrics to determine the meaningfulness of individual meta-symbols are important. Unfortunately, typical evaluations of a recommender system involve calculating the accuracy of its aggregate recommendations against the actions of a historical sample of users or A/B splits of user behavior after using the recommender, with scant attention to the accuracy of an individual recommendation. [17] To address this deficiency, we have developed novel applications of some common metrics used in other fields.

In a Bayesian system, assigned probabilities represent states of knowledge [18], but in principle we could define meta-symbols that represent nothing but noise. A necessary condition of truthfulness and therefore meaning is for the symbol to represent something more than noise. Two metrics are useful: the relative signal and signal-to-noise ratio.

A signal is built up (created) as we record history. If the recorded signal for a specific question response for a specific symbol is similar among the users creating a satisfaction event for that symbol, and yet different from the average for a particular question, then the result is a strong signal. If the recorded signal does not differ from the average for a particular question, then it tells us nothing about the symbol and represents a weak signal.

Of course, different user interests can converge on a single symbol, and different users can like a symbol for different reasons. What we record is an average of all the different underlying signals.

Let's define the signal in the historical record for a single question response associated with a project idea in our

**Figure 1. Sammon mapping of the knowledge space.** (**A**) Sammon mapping of 1,203 base symbols (project ideas) in our recommender system, color coded by 32 different areas of science, enables us to visualize the history of user interactions with the system. Each base symbol is defined by 78 probabilities representing the three possible answers for each of the 26 questions in our science interest survey. Here, Sammon mapping has reduced those 78 dimensions to an estimate of just three arbitrary dimensions that we can plot in a 3-dimensional space. The distance between symbols (in arbitrary units) represents an estimate of their similarity or dissimilarity. Symbols with similar histories will be close together and symbols with dissimilar histories will be far apart. Bubble size scales with the popularity of the symbol in the historical record, larger size meaning that a larger number of students took actions indicating satisfaction with the item.

Symbols can be defined at different levels of abstraction, and this helps us to make sense of the cloud of bubbles in (**A**). (**B**) Sammon mapping of 32 meta-symbols representing select areas of science in our recommender system. Interests that are polar to our users appear on opposite sides of the graph. Colors map to the bubbles in (**A**).

A set of base symbols (project ideas in our recommender) underlie every meta-symbol. (**C**) Shows the base symbols in three areas of science, with a unique color for each area. The three areas overlap, with some areas spreading more than others. Outliers tend to be low popularity (small diameter) symbols, which have noisier underlying data than more popular symbols. In (**D**) we add meta-symbols (charcoal color), defined to represent all symbols in each area. All of the meta-symbols in (**B**) have been defined in the same way. Meta-symbols tend towards the center of the cluster they symbolize.

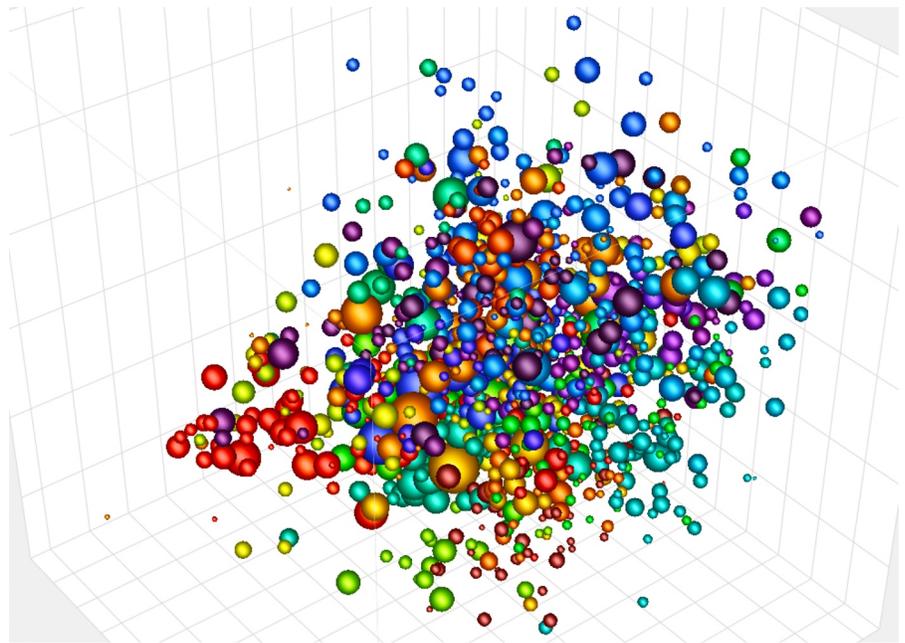

A

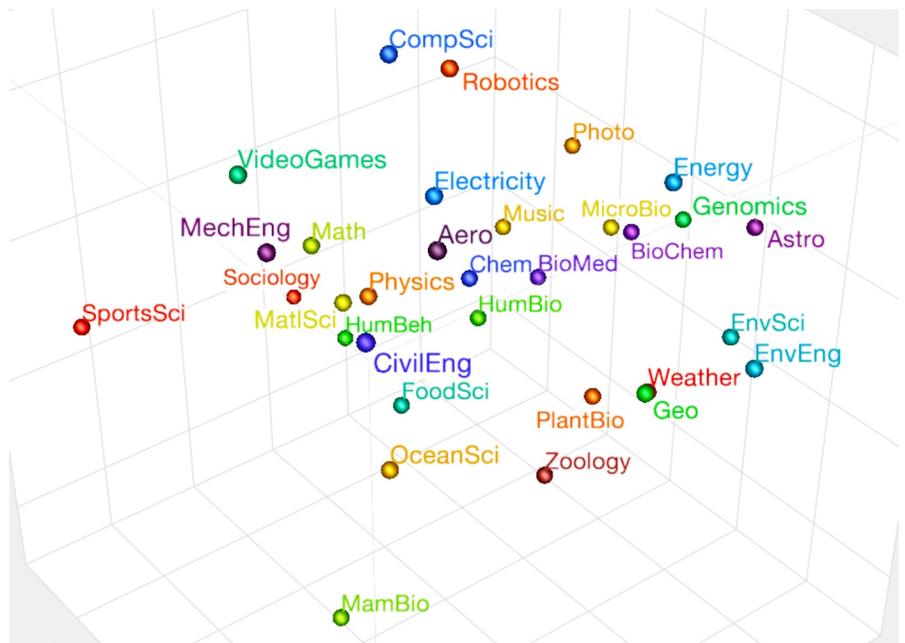

B

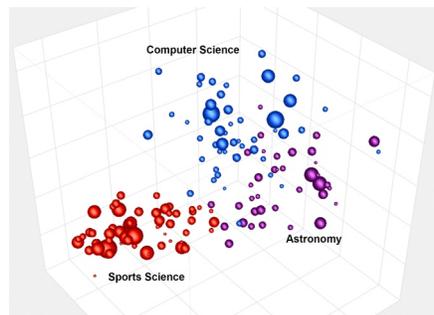

C

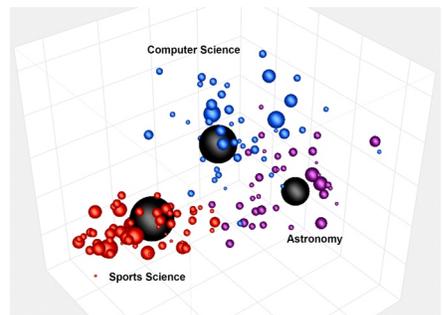

D

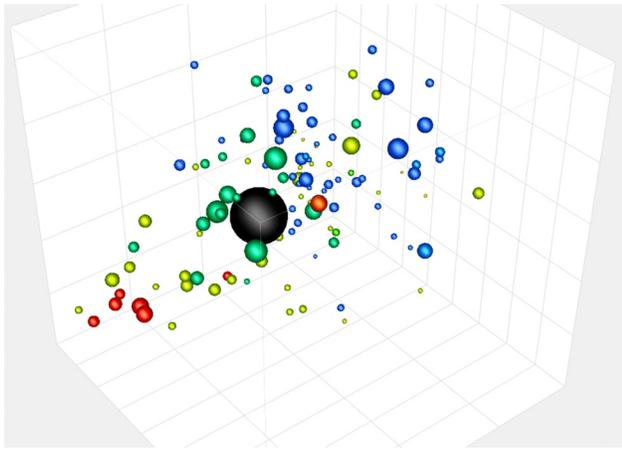 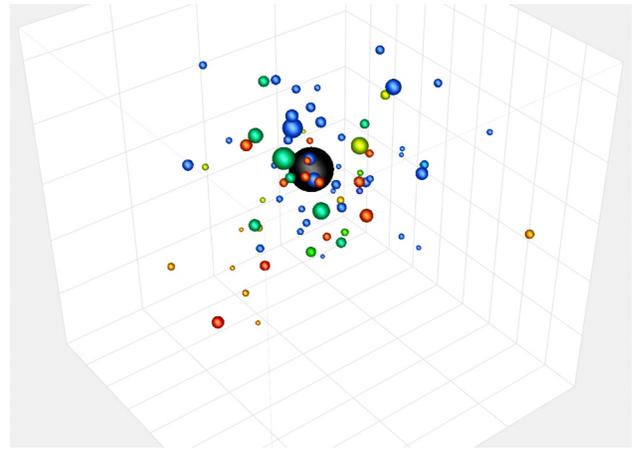

**A**    Computer science and math      **B**    A career in programming

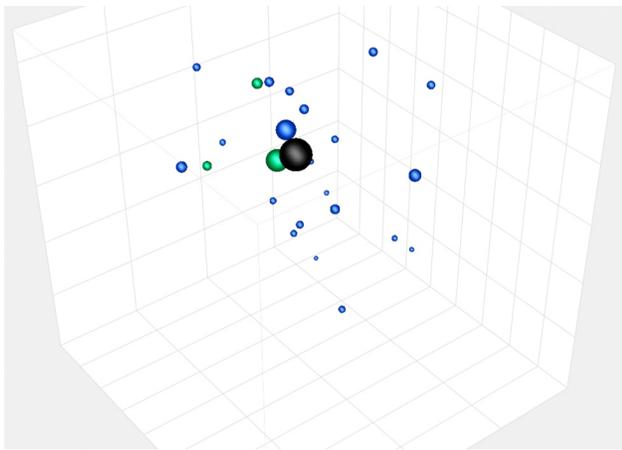 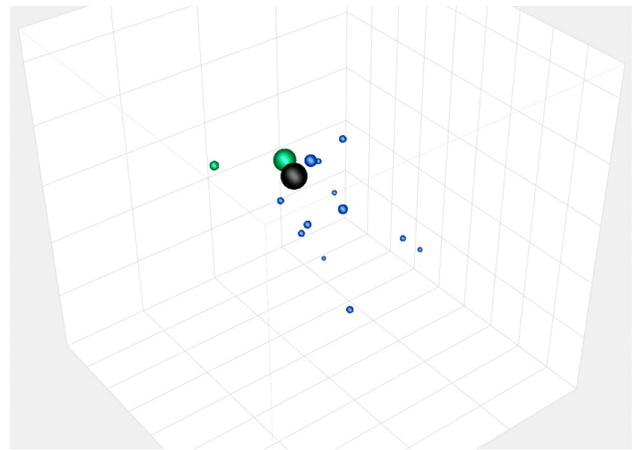

**C**    Computer programming languages      **D**    Scratch programming language recommender. The signal we care about is the difference between the predicted response to a question based on the responses of all users and the actual response to the same question only by those users who expressed an interest in that specific symbol.

**Figure 2. Meta-symbols at different levels of abstraction.** We define meta-symbols with higher levels of abstraction in (**A**) and (**B**) and lower levels of abstraction in (**C**) and (**D**), showing the meta-symbols in charcoal and the underlying base symbols in color (different colors representing different areas of science). The meta-symbols represent (**A**) the over-arching field of computer science and math, (**B**) a career in programming, (**C**) computer programming languages, and (**D**) specifically, the Scratch programming language. In fact, the position of all bubbles is approximate when mapped onto 3 dimensions. Nonetheless, notice how the position of the respective meta-symbol changes with the level of abstraction. In (**A**) and (**B**) base symbols outside the computer science field (colors other than dark blue) pull the meta-symbol in their direction.

For example, in response to the question "Do you enjoy gardening and working with plants?" 45% of all users said "no," 37% said "sometimes," and 18% said "yes." We can multiply these percentages by the number of users who expressed an interest in a specific symbol to calculate the predicted response. By subtracting these predictions from the actual count of user responses, we obtain the signal, the magnitude by which users liking a symbol differ from the overall average.

The signal for symbol $s$, question $q$, and response $r$ is:

$$Signal_{s,q,r} = User\ Response_{s,q,r} - Predicted\ Response_{s,q,r}$$

The total signal for a symbol $s$ is the square root of the sum of the squares of the signals for all survey question responses.

$$Signal_s = \sqrt{\sum_q \sum_r Signal_{s,q,r}^2}$$

**Figure 3. Sammon mapping of users in the knowledge space.** Completing the visualization of our recommender system, in (**A**) we plot users (the outer cloud of 1,000 arbitrarily selected individuals) in the same space as symbols. The coordinates for each user represent their answers to the science interest survey. All 1,203 base symbols (colored bubbles) and more than 2,000 meta-symbols (charcoal bubbles) are in the inner cloud. One of the authors (Hess) is represented by the red bubble, top, just left of center.

From the perspective of the visualization, the user's responses to the science interest survey locate him or her in the knowledge space and the system recommends the closest symbols and meta-symbols to that user. Isolating a single user, (**B**) highlights (red bubbles) the recommender system's top 97 recommendations for one of the authors (Hess, red bubble lower right). All lower-ranking symbols and meta-symbols have a turquoise color, and the orientation of the axes has been adjusted from (**A**) for clarity. Because meta-symbols enable us to identify areas in the knowledge space, we hope that exploration of the asymmetries of the symbol cloud relative to the spherical symmetry of the user cloud will help us identify deficiencies in our content offering and/or the science interest survey.

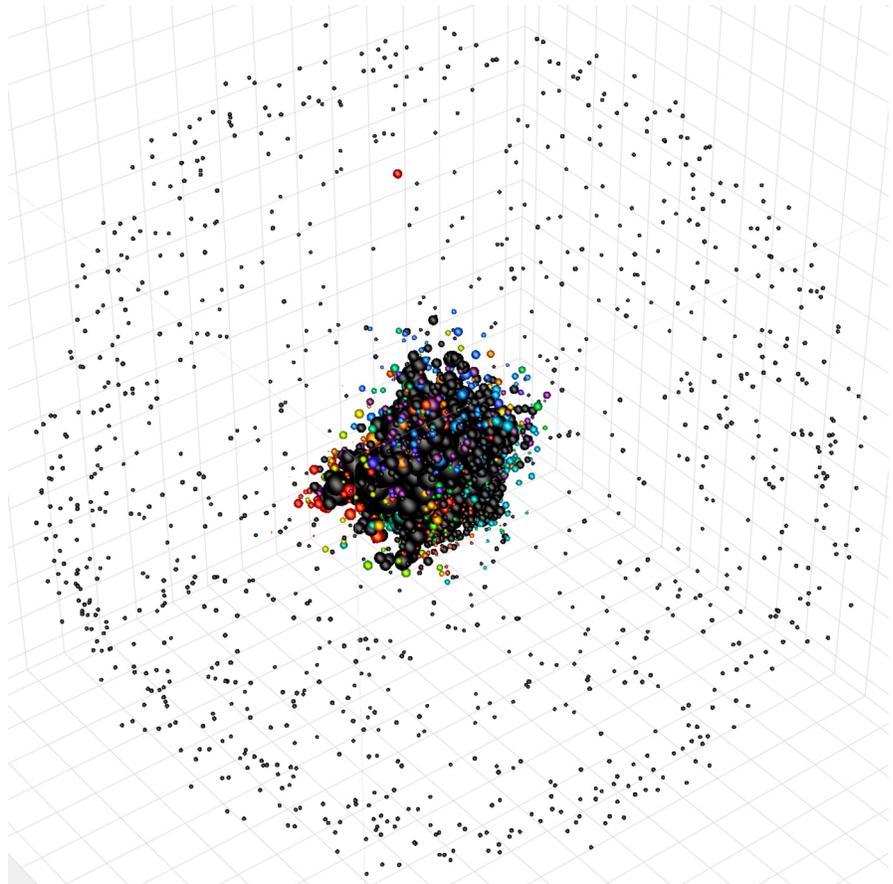

**A**

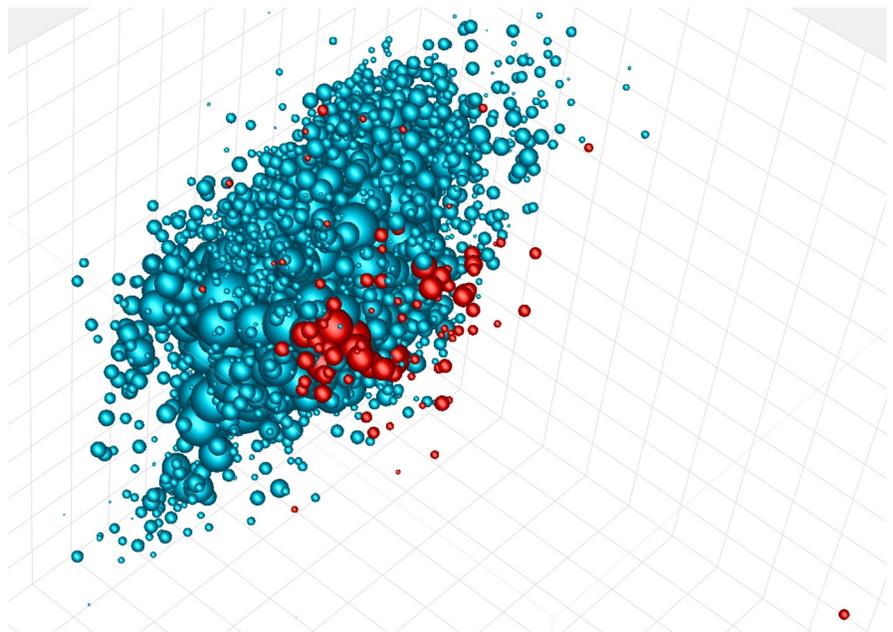

**B**

Calculating the noise in the historical record is more problematic, because there are many sources, not all of them quantifiable at this stage of our research (see Table 3). One source of noise, Shot (Poisson) noise is significant and quantifiable. Shot noise is equal to √N, or in our case the square root of the signal. Because we can only measure some of the noise, we express the signal-to-noise ratio as an inequality, an upper bound. The signal-to-noise ratio ($SNR$) for symbol $s$ is:

$$SNR_s < \frac{Signal_s}{\sqrt{Signal_s}} < \sqrt{Signal_s}$$

---

Noise from errors recording survey responses of the user:
- User does not take survey questions seriously
- User doesn't understand the question or there are multiple interpretations—different interpretations by different users
- Responses to the question change over the integration time of the history

Noise from errors recording user satisfaction events:
- False (positive) satisfaction event
- False (negative) satisfaction event
- Source of interest in a project idea changes over the integration time of the history

---

**Table 3. Summary of some of the known sources of noise in the historical record.**

The SNR for a symbol is driven by the magnitude of the historical record. In fact, SNR tends to grow as one adds related symbols to the definition of a meta-symbol. A tiny signal can have a large SNR, given enough events in the history; hence, we also need to look at the relative signal to better understand whether a symbol has any meaning in the real world.

To measure the relative signal for a single survey question response, we normalize the signal (as calculated above) to the sum of all survey question responses for that question for a single symbol, obtaining a ratio that shows how much each survey question response differs from the overall average. We define the square root of the sum of squares of all 78 survey question responses for each symbol as the total relative signal, a value that is independent of the number of events in the history.

The relative signal for symbol $s$, question $q$, and response $r$ is:

$$Relative\ Signal_{s,q,r} = \frac{User\ Response_{s,q,r} - Predicted\ Response_{s,q,r}}{Total\ User\ Responses_{s,q,r}}$$

The relative signal for symbol $s$ is:

$$Relative\ Signal_s = \sqrt{\sum_q \sum_r Relative\ Signal_{s,q,r}^2}$$

In general, we can have higher confidence in abstract recommendations to a user because of the tendency for abstract meta-symbols to have a higher SNR than base symbols (see Table 4) and that should form the basis for improvements to the user interface presenting the recommendations.

It might seem counterintuitive that meta-symbols defined by extremely large subsets of the universe of symbols in our recommender system do not have signals that regress to the mean. It was counterintuitive to us. However, related symbols (those defining a meaningful meta-symbol) will tend to cluster in one portion of the huge state space. Expanding a definition by adding more related symbols (constructing a meta-symbol with a higher level of abstraction), expands the portion of the state space that the meta-symbol covers. That a meta-symbol does not regress to the mean provides additional confirmation that it has meaning.

Relative signal and SNR can screen out meta-symbols unlikely to be meaningful, but this area needs much more development and the possibilities are numerous. We are currently working on measures of spread. A term with a sharp meaning would ideally have an underlying definition that is sharp (has low spread) as well. Another direction that can sometimes help is to look at the opposite of a term. Is the antonym what one expects? We can easily compute the opposite of any symbol.

**Antonyms: Synthetic Meta-symbols**

In a sense, all meta-symbols are synthetic—some more than others. Meta-symbols comprised of completely random collections of base symbols are more synthetic than those based solely on a traditional area of science. Because we have "digitized" symbols, we can use any number of techniques to synthesize new ones depending on the purpose. We reserve the term "synthetic meta-symbol" for those constructed from a mathematical transformation of the 78-dimensional vector underlying our lingua franca.

We can readily construct meta-symbols that represent an inversion of the signal from another symbol or meta-symbol, their antonym or opposite.

As already mentioned, one reason to compute an antonym is to help understand the term itself. Another reason is to better understand the overall knowledge space. One of our objectives is to identify deficiencies in the breadth of our content offerings. Because these items are not in the database by definition, we need to extrapolate to try to identify them. If we can identify a direction of interest in our knowledge space, then concepts such as "more like this" or "the opposite of that" can become useful to extrapolate outside the existing space.

The user response (history) of the inverse symbol $s$, question $q$, and response $r$ is:

$$Response_{inverse\ s,q,r} = Response_{s,q,r} - 2 \times Signal_{s,q,r}$$

$$Response_{inverse\ s,q,r} = 2 \times Predicted\ Response_{s,q,r} - Response_{s,q,r}$$

| Symbol or Meta-Symbol Category | Avg Relative Signal | Avg SNR | Avg History (# of events) | Number of Examples | Comments |
|---|---|---|---|---|---|
| Base symbols (subset of project ideas, comprising those with limited history) | 0.90 | 3.6 | 29 | 11 | This subset of project ideas behaves as expected with a very low SNR. |
| Base symbols (all project ideas) | 0.74 | 32.4 | 2,158 | 1,204 | Strong relative signal, modest SNR |
| Sub-sub-areas of science, scientific concepts, and scientific techniques (e.g., solubility) | 0.66 | 77.2 | 14,879 | 842 | |
| Sub-area of science (e.g., physical chemistry) | 0.63 | 101.9 | 29,565 | 268 | |
| STEM professions and careers (e.g., chemist)[A] | 0.60 | 131.9 | 50,428 | 152 | |
| Common tools & materials (e.g., pipette) | 0.60 | 78.3 | 22,612 | 250 | This category had a number of meta-symbols that were under-defined and should be removed in a production version of the recommender. |
| Area of science (e.g., chemistry) | 0.51 | 255.8 | 177,651 | 32 | |
| Interdisciplinary areas of science (e.g., chemistry + computer science) | 0.38 | 299.3 | 310,176 | 630 | |
| Major area of science (e.g., physical science) | 0.42 | 384.6 | 473,051 | 6 | A couple of these areas are over-defined, including peripherally related base symbols that should be removed in a production version of the recommender. |
| Aggregated meta-symbol comprising all base symbols | 0.00 | 5.4 | 2,596,097 | 1 | Relative signal zero by definition |
| Meta-symbols comprising random selection of approx. 100 base symbols each | 0.12 | 151.1 | 191,397 | 50 | Very weak relative signal as it should be, and SNR below other symbols w/similar history |

[A] For the native career recommender described in the article, the corresponding metrics are: average relative signal = 0.56, average SNR = 13.8, and average history = 437.

**Table 4. Comparison of relative signal and SNR by category of symbol and meta-symbol.** Driven by the magnitude of the underlying history, there is a tendency for meta-symbols with higher levels of abstraction to have higher SNR. When that is combined with a strong relative signal, we can have strong confidence in the meaning of the symbol. Symbols with a limited history can have strong relative signals, but the SNR indicates we should not put faith in them, whereas randomly defined meta-symbols can have a high SNR, but the small relative signal is exactly what we would expect. Neither of those categories of meta-symbols reliably conveys meaning about the real world. Note: These results are unfiltered; we have not yet excluded items that fail to meet reasonable trade-offs for relative signal or SNR. For example, we have items that appear to be over-defined (comprised of too many base symbols, some of which are peripheral to the definition) and under-defined (too few base symbols to form a meaningful definition).

In some cases, the probabilities for inverse symbols can take on a negative value, which is not a problem for our purposes, but we will call them pseudo-probabilities to avoid confusion. The pseudo-probability of inverse of symbol $s$, question $q$, and response $r$ is:

$$PseudoProbability_{inverse\ s,q,r} = \frac{Response_{inverse\ s,q,r}}{Total\ Responses_{s,q,r}}$$

$$PseudoProbability_{inverse\ s,q,r} = 2 \times \frac{Predicted\ Response_{s,q,r}}{Total\ Responses_{s,q,r}} - Probability_{s,q,r}$$

Figure 5 shows meta-symbols for 32 areas of science with their corresponding inverse meta-symbols.

**Results and Discussion**

For our end users, we want to be able to communicate something richer than a ranked list of project ideas, and the new symbolic capabilities make that possible. The lingua franca can describe the areas of science that a student would be most interested in, starting at a very general level (e.g., physical vs. life science) and then diving down to narrower and narrower specialties, including novel interdisciplinary areas.

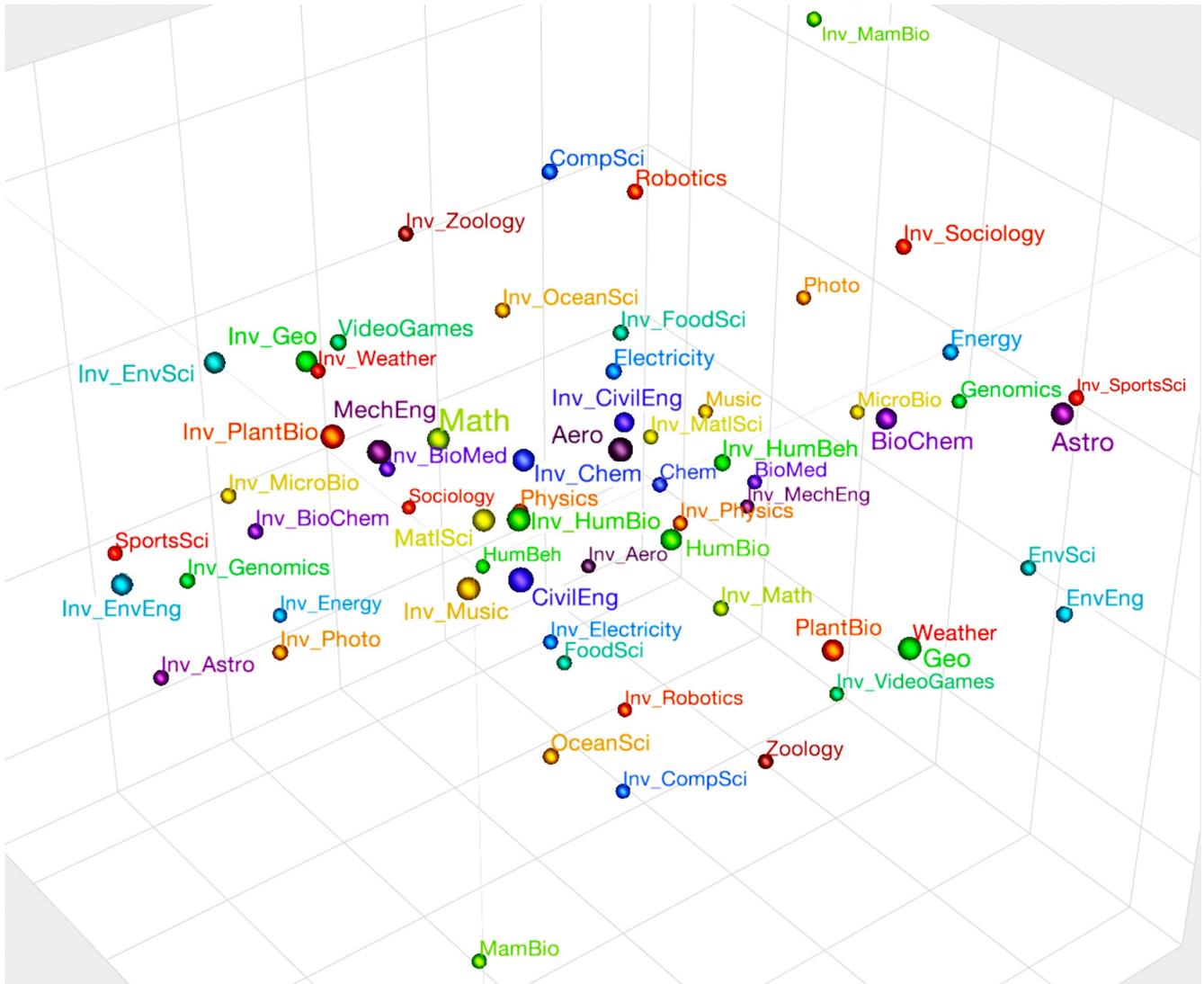

**Figure 5. Meta-symbols for inverse areas of science.** Sammon mapping of 32 areas of science with their inverses, which occupy opposite positions in the knowledge space. Because the cloud of area of science meta-symbols is oblate and lacks symmetry in some directions, the inverse symbols are all that is available to describe some areas of the knowledge space. In other cases, meta-symbols and inverse meta-symbols are next to each other, leading to some thought-provoking comparisons. Note that this knowledge space is defined by the history of approximately 1.5 million K-12 students.

Importantly, over a very broad range, the more abstract (higher level) the meta-symbol, the higher its SNR, making these some of the system's most confident recommendations. The project ideas, instead of comprising the only output, can become examples of the more general descriptions. We are currently prototyping a user interface to make this new capability available to the public.

As part of implementing this new capability, there are a number of engineering trade-offs that remain to be determined, such as the appropriate thresholds for relative signal and SNR.

As designers and maintainers of the system, we want better tools to understand how changes to the system impact the quality of its recommendations. For example, if we add or subtract questions from our science interest survey, what is the relative impact on the ranking of different areas of science? Do the various disciplines rank similarly to each other after the change, or do the changes result in a skew towards one area or another? Combining the lingua franca with Spearman Rank Correlations of alternative sets of science interest survey questions has proven to be a powerful analytical tool, and we have just begun a thorough analysis of our system.

Importantly, we also want to identify deficiencies in the breadth of our content offerings. Can the system identify areas of science that are intrinsically interesting to our audience that we do not currently offer? This is the most challenging of our objectives and the jury is still out as to how helpful the new capabilities will be.

There have been a number of unanticipated bonus results from this work. Perhaps most important is the ability for a lingua franca to address the cold-start problem. As we have demonstrated with careers, by defining the new area for recommendations recursively in terms of the existing area where the system already has data to make recommendations, the resulting lingua franca can bridge the system into the new area where it has yet to obtain data. On a micro level, this same technique can be used when inserting a new item into the existing area of recommendations.

Sammon mapping began as a way to visually validate ideas and communicate with others; however, the visualizations ultimately stimulated us to ask new questions that we had not considered. For example, Sammon mapping highlighted a relationship between relative signal and the prior probability. If we ignore prior probability, outliers with a limited history and high relative signal move to the surface of the symbol/meta-symbol cloud, closer to users and more likely to be recommended. While using the full prior probability turns recommendations into a popularity list, ignoring prior probability is just as bad. This reinforces the appropriate weighting for the prior probability as a key design trade-off. Sammon mapping also made us aware of the need to measure and analyze the spread in the knowledge space of the underlying base symbols in a definition.

Earlier, we summarized our objectives as a desire to understand how deep our recommender's understanding goes. As others have found, what we learned is that the historical data was extraordinarily rich in ways that we did not always anticipate. [19] The system knows a lot!

**Knowledge Representation**

In our system, symbols and meta-symbols are not identified with a single survey response (a single input). While some responses carry more weight than others, every input plays a role in ranking every symbol, meaning that knowledge is widely distributed throughout the system.

The same underlying history, probed with meta-symbols having different definitions, can unlock entirely different kinds of knowledge. It is like looking at an object through different lenses. As demonstrated, by defining the proper meta-symbols, a history about interest in science projects is found to contain knowledge about STEM careers. Within the scope of the knowledge space, the process is one of asking new, carefully formed questions, not one of building a new history or retraining the existing one.

As we add more and more meta-symbols to the system, the output of the system could become overwhelming to the user. We can attenuate this problem, increasing the selectivity by filtering out unwanted results, focusing on those the user cares about at that instant. In our case, show the user project ideas when working on a science project and STEM careers when exploring what to do with the user's life.

So, by tuning our frame of reference with novel meta-symbols and a dynamic focus of attention, we can extract different chunks of knowledge from the same underlying history.

The historical record comprising counts of events is roughly equivalent to the firing rates of a collection of neurons. The time interval is a constant, so the counts of events are mathematically the same as a rate. And, the system of inputs, symbols, and meta-symbols can be thought of as an interconnected network. This raises the question of whether knowledge representation in nature follows a similar model.

**Future Work**

As we work on a prototype end-user interface for the lingua franca, we continue to generate new ideas for extensions of the technique.

Appropriately constructed meta-symbols could be powerful tools for analyzing subsets of users to better understand their needs. And, we could aggregate users into meta-users (analogous to the process for generating meta-symbols) as a way to make group recommendations.

While this paper has focused on the lingua franca output from the system, it can also be used to provide input, enabling a new means for users to communicate their preferences. We are actively working on this capability.

We're also interested in other applications of this technique. To construct a similar lingua franca requires the existence of:

1. Individual items that symbolize additional features
2. A history of choices, actions, or outcomes that are correlated with features of the items involved
3. The means to aggregate the history of a set of items
4. The means to rank items based on current choices, actions, or outcomes

It is not a requirement to use a Naïve Bayes algorithm. These are not difficult requirements to meet, and we are excited about the possibilities.